# Divanillin-based aromatic amines: synthesis and application as curing agents for bio-based epoxy thermosets

**Etienne Savonnet[1,2,3], Cedric Le Coz[1,2], Etienne Grau[1,2], Stéphane Grelier[1,2], Brigitte Defoort[3], Henri Cramail[1,2]\***

[1]Univ. Bordeaux, Laboratoire de Chimie des Polymères Organiques, UMR 5629, 33607 Pessac Cedex, France

[2]Centre National de la Recherche Scientifique, Laboratoire de Chimie des Polymères Organiques, UMR 5629, 33607 Pessac Cedex, France

[3]ArianeGroup, Rue du général Niox, St-Médard-en-Jalles, 33160, France

**\* Correspondence:**
cramail@enscbp.fr

## Abstract

New bio-based diamines were successfully synthesized from vanillin and fully characterized. These amines, methylated divanillylamine (MDVA) and 3,4-dimethoxydianiline (DMAN), were then used as curing agent with epoxy monomers. Epoxy thermosets obtained from these new bio-based amines exhibited promising thermomechanical properties in terms of glass transition temperature and char residue. These latter could be valuable alternative to conventional amine hardener.

## 1    Introduction

Epoxy thermosets are the products of the reaction between epoxy-based monomers or prepolymers with curing agents. The cross-linking agents are significant in terms of mass fraction as the latter can represent up to 50% of the formulations. Many efforts are currently done to found bio-based alternatives to traditional epoxy monomers derived from fossil resources and in particular bisphenol-A.(Kumar2018) With similar objectives, some studies report the synthesis of bio-based anhydride- or acid-type curing agents from renewable resources, such as vegetable oils,(Roudsari2014) rosin,(Liu2009, Qin2014a, Wang2008, Wang2009) terpens,(Takahashi2008) tannins (Pizzi2008, Shibata2010) or lignins.(Qin2014b)

Moreover aliphatic amines curing agents have also been reported from molecules originating from biomasses such as terpenes,(Garrison2016, Keim1981, Keim1983) lignin,(Fache2014) cardanol (Shingte2017, Thiyagarajan2011) derived from cashew oil or vegetable oil.(Hibert2016, Samanta2016, Stemmelen2011) Very recently amines have been synthesized from vanillin (Mora2018) by $NH_3$ addition on diglycidilated vanillin alcohol. When cured with DGEBA, Tg of 72°C is obtained.

Despite all these works, the development of bio-based fully aromatic amine-type curing-agents leading to high Tg epoxy resins is still an unmet challenge. In this way, there is a growing interest to find bio-based reactive amines leading to epoxy thermosets with high thermomechanical properties.

Vanillin is a very interesting candidate because it is one of the nonhazardous aromatic compounds industrially available from biomass.( Pinto2012) From vanillin, we developed an efficient C−C coupling through enzymatic catalysis leading to divanillin.(Kobayashi2009, Llevot2015, Llevot2016) More recently, we have developed a palette of epoxy monomers derived from divanillin, which demonstrated to be valuable and realistic alternative to DGEBA-based epoxy thermosets.(Savonnet2018) With the same strategy, this article presents two synthetic pathways to prepare primary aromatic amines from divanillin as starting material. These polyfunctional amines were then used as curing agents for the synthesis of epoxy thermosets.

## 2 Synthesis of bio-based amines from vanillin derivatives

Herein, two synthetic pathways have been identified for the synthesis of bio-based aromatic amines through the reduction of oxime and the oxidative rearrangement acyl azide moieties leading to a bis-benzylamine and bis-aniline moieties respectively.

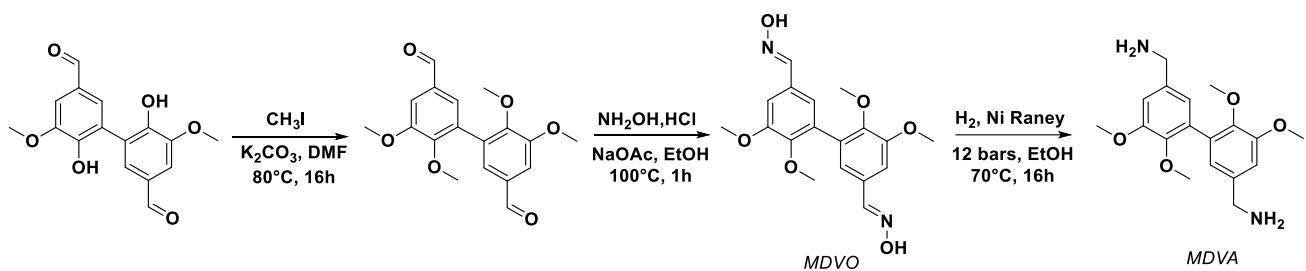

**Scheme 1: Synthesis of methylated divanillylamine from divanillin**

The divanillin displays two aldehyde functions which can undergo chemical reaction to get divanillyloxime. A synthetic pathway adapted from literature has been developed and is depicted on Scheme 1.(Fache2015, Liu2011) First, the alkylation of phenol moiety was performed in the presence of iodomethane and a weak base leading to methylated divanillin in a very good yield (>95%). Then, the oximation step consisted in the reaction of aldehyde functions with hydroxylamine hydrochloride in the presence of sodium acetate to yield methylated divanillyloxime (MDVO) (>95%). The structure of this intermediate was confirmed by $^1$H and $^{13}$C NMR spectroscopy (Figure S1). The appearance of a signal at 3.67 ppm was attributed to the methylated phenols and new signals at 8.10 and 11.13 ppm were attributed to the oxime moieties of methylated divanillyloxime (MDVO). Finally, the so formed oxime was then reduced into methylated divanillylamine (MDVA) by hydrogenation. The reaction was performed during 16 h at 70 °C under 12 bars of pressure in the presence of Nickel Raney in ethanol. Reduction of the oxime yielded a pale orange solid with a melting temperature of 69 °C (DSC). $^1$H NMR spectrum of this orange solid demonstrated the disappearance of oxime signals at 8.10 and 11.13 ppm and the appearance of a new signal at 3.63 ppm, corresponding to the benzylic-protons of the amine (Figure 1). $^{13}$C NMR spectrum exhibited also a shift of the alpha-carbon of the oxime from 147.81 to 51.62 ppm, which confirmed the reduction of oxime moieties. The attribution of the signals was also confirmed by HSQC NMR spectroscopy. In addition, the MDVA was also characterized by FTIR. Figure S2a shows the infra-red spectra of MDVO and MDVA from which it is possible to observe the decrease of O-H stretching signal at 3227 cm$^{-1}$ and the disappearance of N-O stretching at 945 cm$^{-1}$, corresponding to the oxime moieties. Besides, ninhydrin test,(Friedman2004) which reveals amino groups, was realized and confirmed the presence of amine moieties on MDVA (Figure S2b).

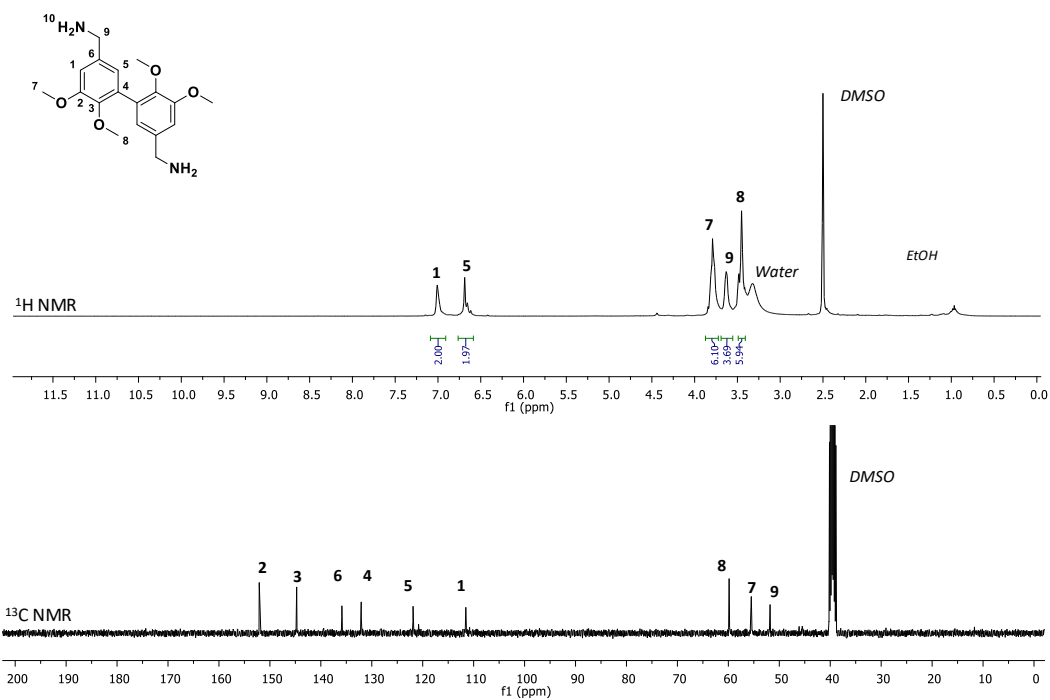

**Figure 1: $^1$H and $^{13}$C NMR spectroscopy of methylated divanillylamine (MDVA) in DMSO-d6**

The second pathway is based on the Curtius rearrangement. It involves the synthesis of an acyl azide intermediate. Usually carboxylic acids are precursors of acyl azides. As previously described,(Kobayashi2009, Llevot2015, Llevot2016) methyl vanillate can easily be dimerised into the corresponding dimer and hydrolysed into the diacid. From divanillic acid, the sequential synthetic pathway is summarised in Scheme 2. All the intermediates were obtained without any further purification steps, except mentioned. $^1$H and $^{13}$C NMR spectroscopies were performed to confirm the structure of the synthesized products. Methyl divanillate was first alkylated using the same procedure described previously and obtained in a good yield (>80%). Thereafter, the hydrolysis of the methylated diester with sodium hydroxide yielded the diacid (>90%). The disappearance of the methyl ester protons at 3.38 ppm and the appearance of acid proton signals of carboxylic acid at 9.20 ppm were attributed to the formation of the methylated divanillic acid. Methylated divanillic acid was then converted into acyl azide in a two-step reaction. Ethyl chloroformate was first reacted with the acid to form in situ an acyl chloride and sodium azide was then added to the mixture yielded the corresponding acyl azide (>60%). Finally, di-isocyanate was obtained in good yield (>80%) by simply heating the azide compound in dry toluene. The synthesis of these latter compounds was confirmed by $^1$H NMR spectroscopy (Figure S3).

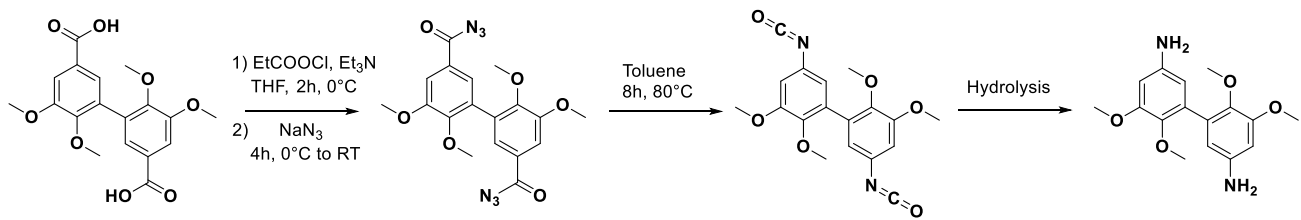

**Scheme 2: Synthesis of 3,4-dimethoxydianiline (DMAN) from methylated divanillic acid**

In addition, acyl azide structure was confirmed by the disappearance of carboxylic signal at 9.20 ppm and finally the shift of the aromatic protons from 7.61 and 7.55 to 6.65 and 6.58 ppm was attributed to the so-formed di-isocyanate. These compounds were also characterised by FTIR spectroscopy (Figure S4). Infrared spectra exhibited characteristic signal wavelengths of carboxylic acid at 1720 cm$^{-1}$ (-C=O) and 2500- 3300 cm$^{-1}$ (O-H), acyl azide at 2140 cm$^{-1}$ (-N$_3$) and isocyanate at 2278 cm$^{-1}$ (-N=C=O).

Finally, 3,4-dimethoxydianiline (DMAN) was recovered by hydrolysis in basic conditions of the corresponding di-isocyanate. After extraction with ethyl acetate and washing with water a mixture of brown and white solids was obtained. However, $^1$H NMR spectroscopy of the reaction mixture revealed the presence of a by-product (Figure S5). These additional signals could be attributed to the formation of ureas, resulting of the side reaction of amine with the isocyanate or to the oxidation of the amines. Nevertheless, a small fraction was isolated as a white solid corresponding to the DMAN and characterized by $^1$H and $^{13}$C NMR spectroscopy (Figure 2). $^1$H NMR spectroscopy displayed protons signals of aromatic rings, amine moieties and alkylated hydroxyl groups at 6.23, 5.90, 4.79, 3.72 and 3.38 ppm, respectively. In addition, the disappearance of the C9 carbon signals of isocyanate at 124.71 ppm, confirmed the obtention of the targeted diamine. DMAN was also characterised by FTIR spectroscopy (Figure S6). The following spectrum confirmed the absence of the isocyanate bands (-N=C=O) at 2278 cm$^{-1}$ and the appearance of amino groups signals at 1610, 3240, 3370 and 3471 cm$^{-1}$. In conclusion, the Curtius rearrangement permitted to synthesize the 3,4-dimethoxydianiline from the acyl azide derived from the methyl divanillate. However, the last step of the synthesis was tedious and only a small amount of the desired product was identified as the diamine. Further investigations and optimisations in the hydrolysis step of isocyanate are required to recover the bio-based diamine in a better yield.

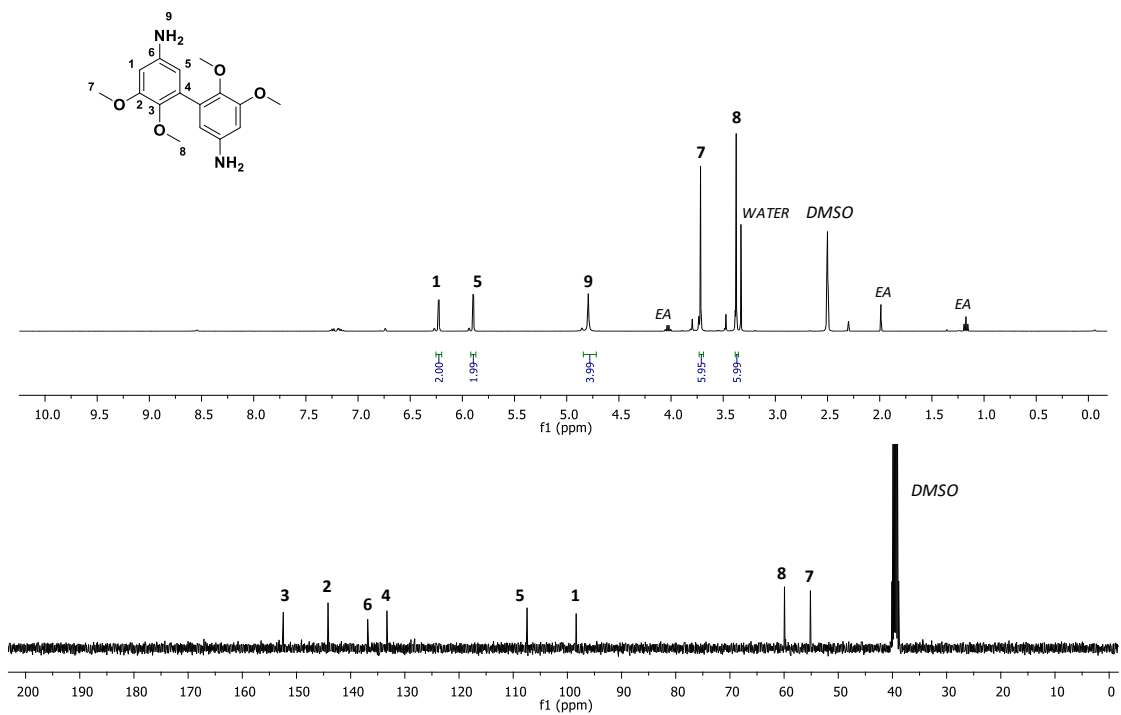

**Figure 2: $^1$H and $^{13}$C NMR spectroscopy of 3,4-dimethoxydianiline (DMAN) in DMSO-d6**

# 3    Synthesis of fully vanillin-based epoxy thermosets

The previously synthesized di-amines, i.e. methylated divanillylamine (MDVA) and 3,4-dimethoxydianiline (DMAN) were used as curing agents for the synthesis of epoxy thermosets.

The DGEBA epoxy monomer was stirred vigorously with a stoichiometric amount of MDVA and the thermomechanical properties of the so-formed network were characterized by DSC and TGA. Bio-based thermoset was then compared with DGEBA/IPDA epoxy system. Results are summarised in the Table S1. Unfortunately, the characterization of bio-based epoxy networks, using DSC, showed a weak exothermic peak ($\Delta H=105$ J/g for DGEBA/MDVA vs 430 J/G for DGEBA/IPDA), corresponding to the reaction of amine with epoxy. In addition, no clear glass transition temperature was observed on this network (Figure S8). One hypothesis to explain this result could be the poor homogeneity of the mixture between DGEBA and the solid MDVA, which impaired the stoichiometric ratio. Indeed, on the DSC, one can observe the melting of MDVA at 40°C.

DGEBA epoxy monomer was also cured with bio-based DMAN. However, in view of the small amount recovered, crude DMAN was used for the curing reaction. Thermomechanical properties of network were then characterized and compared with DGEBA networks cured with DDS. Results are summarised in the Table 1 and Figure S8b. In comparison with conventional amine hardener DDS, epoxy network cured with DMAN exhibited similar properties. Indeed, glass transition temperature of thermoset cured with DMAN displayed a value 30 °C below the networks obtained with DDS i.e. 176°C. This difference can be explained by the presence of by-product compounds, which could impair the stoichiometric ratio. Interestingly, despite the presence of the by-product, the bio-based amine enabled to increase the char yield residue up to 28% (Figure S9). This statement could be explained by the C-C bonding between the two aromatic rings of DMAN, which could favour the formation of char and thus increased the residual content. Another important difference between DDS- and DMAN-based epoxy thermoset is the reactivity of DDS

**Table 1: Thermomechanical properties of DGEBA and TetraGEDVA cured with DDS and DMAN**

| Epoxy/ hardener | $T_{onset}$ (°C) | $T_{Exo}$ (°C) | $\Delta H$ (J.g$^{-1}$) | $\Delta H$ (kJ.mol$^{-1}$) | Tg (°C) | Char$_{900}$ (%) |
|---|---|---|---|---|---|---|
| DGEBA/DDS | 184 | 226 | 355 | 83 | 204 | 16 |
| DGEBA/DMAN | 99 | 153 | 393 | 95 | 176 | 28 |
| TetraGEDVA/ DDS | 165 | 208 | 459 | 89 | - | 48 |
| TetraGEDVA/ DMAN | 110 | 193 | 183 | 39 | 212 | 48 |

At last, a bio-based polyglycidylether, tetraglycidylether of divanillyl alcohol (TetraGEDVA),(Savonnet2018) was cured with crude DMAN. In this way, a fully bio-based epoxy thermoset was successfully obtained. Bio-based epoxy networks exhibited promising thermomechanical properties with glass transition temperature of 212 °C and char residue of 48% (Figure S9). However, enthalpy is much lower than the enthalpy of the reaction between TetraGEDVA and DDS. Once again, the presence of undefined compounds in crude DMAN and the inhomogeneity of the mixture could explain this difference

## 4    Conclusion

In summary, the synthesis of bio-based curing agent from divanillin derivatives was investigated. Two routes were chosen to achieve the synthesis of diamines. The first one consisted in the reduction of divanillyloxime obtained from divanillin. The synthesis yielded to methylated divanillylamine (MDVA) and the thermomechanical properties of epoxy precursors cured with MDVA were investigated. DGEBA epoxy prepolymers were thus cured with MDVA and the networks obtained were compared with the conventional DGEBA/IPDA system. Unfortunately, the characterization by DSC showed a weak exothermic peak, corresponding to the reaction of amine with epoxy, but no clear glass transition temperature was observed on this network.

Then, the 3,4-dimethoxyaniline (DMAN) was synthesized using the oxidative rearrangement of Curtius. From methyl divanillate and through the synthesis of acyl azide and isocyanate intermediates, the hydrolysis of this latter yielded to 3,4-dimethoxyaniline (DMAN). However, the hydrolysis step was tedious and the corresponding amine failed to be isolated efficiently. Nevertheless, crude DMAN was used as curing agent in the polyaddition reaction with DGEBA. Interestingly, the so-formed semi bio-based thermoset exhibited a glass transition temperature of 176 °C against 204 °C for the conventional DGEBA/DDS system. Moreover, the DMAN permitted to double the char content of the network comparing to DGEBA/DDS network cured in the same conditions. Moreover, a fully bio-based epoxy thermoset was obtained by curing tetraglycidylether of divanillyl alcohol (TetraGEDVA) with DMAN. The thermomechanical properties obtained were promising as the network exhibited a glass transition temperature of 212 °C and a char residue of 48%.

Finally, new amine-type curing agent from vanillin-based starting material was successfully attempted and characterized. Nevertheless, further optimisations in the synthetic pathways described are necessary in order to have a better appreciation of the potential of these new bio- based aromatic diamines.

## 5    Conflict of Interest

The authors declare that there are no conflicts of interest.

## 6    Author Contributions

ES performed the synthesis of the chemicals and performed the polymerization experiments. CC performed the thermo-mechanical analysis. EG, SG, BD and HC contributed equally to the supervision of the project.

## 7 Acknowledgments

The authors thank Gerard Dimier for the DSC analyses. This study was financially supported by ArianeGroup and ANRT. The authors also thank Equipex Xyloforest ANR-10-EQPX-16 for flash chromatography.

## 8 Reference styles

# Divanillin-based aromatic amine precursors as curing agents for bio-based epoxy thermosets

*Etienne Savonnet[a,b,c], Cédric Le Coz[a,b], Etienne Grau[a,b], Stéphane Grelier[a,b], Brigitte Defoort[c] and Henri Cramail\*[a,b]*

a.   Univ. Bordeaux, Laboratoire de Chimie des Polymères Organiques, UMR 5629, 33607 Pessac Cedex, France. E-mail: cramail@enscbp.fr

b.   Centre National de la Recherche Scientifique, Laboratoire de Chimie des Polymères Organiques, UMR 5629, 33607 Pessac Cedex, France.

c.   ArianeGroup, Rue du général Niox, St-Médard-en-Jalles, 33160, France.\*E-Mail: cramail@enscbp.fr; web: www.lcpo.fr

This supporting information contains 9 figures over 13 pages.

## Experimental Section:

## Materials

Laccase from Trametes versicolor, benzyltriethylammonium chloride (99 %), bisphenol-A diglycidylether (D.E.R ™ 332), hydroxylamine hydrochloride (99%), tetraethylammonium bromide (98 %), 4,4'-Diaminophenyl sulfone (97 %), hydrochloric acid (37,5 %), iodomethane (99 %), potassium hydrogen phthalate (99 %), ethylchloroformate (97 %) *meta*-chlorobenzoic acid (77 %) and sodium percarbonate (available $H_2O_2$ 20-30%) were purchased from Sigma-Aldrich.

Epichlorohydrin (99 %), eugenol (99 %), hydroxytosyloxyiodobenzene (97%) and isophorone diamine (97 %) were purchased from TCI.

Vanillin (99 %), methyl vanillate (99 %), perchloric acid solution (0.1 N), sodium borohydride (99 %) and isophorone diamine (99 %) were purchased from Acros.

Potassium carbonate (99 %), triethylamine (99 %), potassium hydroxide (pellet) and sodium hydroxide (pellet) were purchased from Fisher.

TetraGEDVA was synthesized according to reported procedure.[1]

All products and solvents (reagent grade) were used as received, unless mentioned explicitly.

### Procedure for dimerization of phenols

A solution of vanillin (5mmol) in acetone (20 mL) was added to NaOAc buffer (180 mL, 0.1 M, pH 5.0). The solution was saturated in $O_2$ for 5 min. Laccase from *Trametes versicolor* (20 U, 12.4 mg) was added and the reaction was stirred at room temperature for 24 h. The precipitate was filtered off the solution and the product dried overnight at 80 °C under vacuum. Yield: 90%

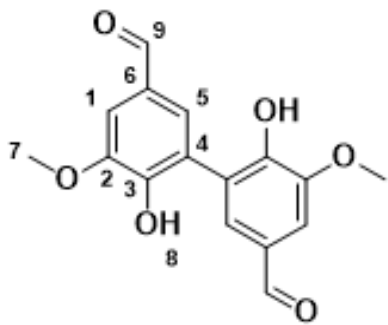

[1]H NMR (400 MHz, DMSO, δ (ppm)): δ 9.69 (s, $H_9$), 7.57 (d, $H_1$), 7.16 (d, $H_5$), 3.76 (s, $H_7$).

[13]C NMR (400 MHz, DMSO, δ (ppm)): δ 191.62 (s, $C_9$), 150.88 (s, $C_3$), 148.61 (s, $C_2$), 128.64 (s, $C_6$), 128.21 (s, $C_4$), 125.02 (s, $C_5$), 109.6 (s, $C_1$), 56.25 ($C_7$).

The same procedure was applied for methyl vanillate. Yield: 90%

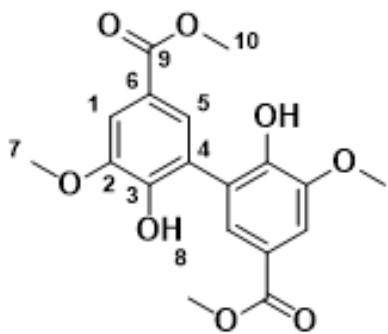

$^1$H NMR (400 MHz, DMSO, δ (ppm)): δ 9.51 (s, $H_8$), 7.46 (d, $H_1$), 7.45 (d, $H_5$), 3.90 (s, $H_7$), 3.80 (s, $H_{10}$).

$^{13}$C NMR (400 MHz, DMSO, δ (ppm)): δ 166.09 (s, $C_9$), 148.88 (s, $C_3$), 147.47 (s, $C_2$), 125.40 (s, $C_5$), 124.36 (s, $C_6$), 119.48 (s, $C_4$), 110.92 (s, $C_1$), 56.01 (s, $C_7$), 51.79 (s, $C_{10}$).

## *Procedure for methylation*

26 mmol of divanillin and 15,2 g of potassium carbonate (110 mmol) were dissolved in 120 mL of DMF. 9,6 ml of iodomethane (158 mmol) were slowly added to the mixture. After 15 h at 80 °C, mixture was filtered and the resulting solution poured into cold water. The methylated compound, which precipitated was filtered off and dried under vacuum. Yield: 80%.

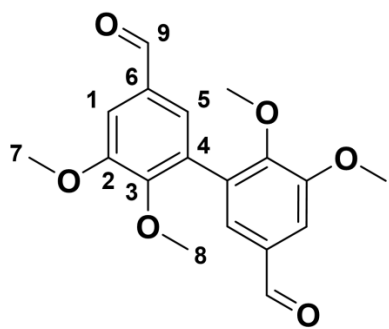

$^1$H NMR (400 MHz, DMSO, δ (ppm)): δ 9.94 (d, $H_9$), 7.58 (d, $H_1$), 7.45 (d, $H_5$), 3.95 (s, $H_7$), 3.67 (s, $H_8$).

$^{13}$C NMR (400 MHz, DMSO, δ (ppm)): δ 191.76 (s, $C_9$), 152.88 (s, $C_2$), 151.52 (s, $C_3$), 131.81 (s, $C_6$), 131.56 (s, $C_4$), 126.09 (s, $C_5$), 111.36 (s, $C_1$), 60.43 (s, $C_8$), 55.99 (s, $C_7$).

The same procedure was applied for methyl divanillate. Yield: 80%

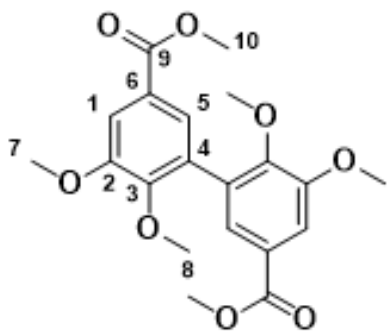

$^1$H NMR (400 MHz, CDCl3, δ (ppm)): δ 7.62 (d, $H_1$), 7.57 (d, $H_5$), 3.95 (s, $H_7$), 3.88 (s, $H_{10}$), 3.71 (s, $H_8$).

$^{13}$C NMR (400 MHz, CDCl3, δ (ppm)): δ 166.79 (s, $C_9$), 152.54 (s, $C_3$), 151.06 (s, $C_2$), 131.87 (s, $C_6$), 125.30 (s, $C_4$), 125.10 (s, $C_5$), 113.09 (s, $C_1$), 60.95 (s, $C_8$), 56.14 (s, $C_7$), 52.25 ($C_{10}$).

### *Procedure for ester hydrolysis*

10 mmol of methylated diester were dissolved in 30 mL of methanol. 3g of sodium hydroxide (75 mmol) were slowly added to the mixture and warmed to reflux during 4h. After cooling at room temperature, the solution is acidified with HCl to pH=3. The precipitate was filtered off and the product dried overnight at 80 °C under vacuum. Yield: 90%

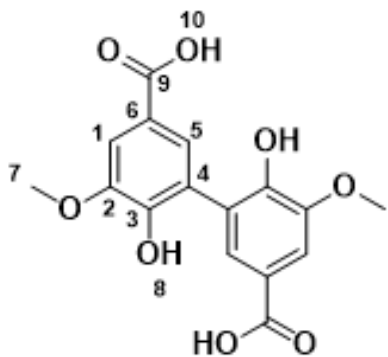

$^1$H NMR (400 MHz, DMSO, δ (ppm)): δ 9.39 (s, $H_8$), 7.45 (d, $H_1$), 7.41 (d, $H_5$), 3.89 (s, $H_7$).

$^{13}$C NMR (400 MHz, DMSO, δ (ppm)): δ 167.18 (s, $C_9$), 148.36 (s, $C_3$), 147.22 (s, $C_2$), 125.44 (s, $C_6$), 124.19 (s, $C_4$), 120.44 (s, $C_5$), 111.05 (s, $C_1$), 55.89 (s, $C_7$).

### *Procedure for oximation*

1 g of hydroxylamine hydrochloride (7 mmol) and 2 g of sodium acetate (12 mmol) were solubilised in 20 mL of ethanol (+4 mL of water). 2 g of divanillin or methylated divanillin (6 mmol) were added to the mixture. After 2 h of magnetic stirring at 100 °C, the product is extracted with dichloromethane and washed with water and dried under vacuum. Yield: 85%

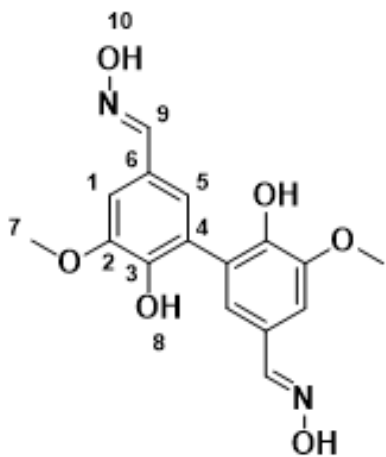

$^1$H NMR (400MHz, DMSO, δ (ppm)): δ 10.83 (s, $H_{10}$), 8.84 (s, $H_8$), 8.02 (s, $H_9$), 7.18 (s, $H_4$), 6.94 (s, $H_5$), 3.85 (s, $H_7$).

$^{13}$C NMR (400MHz, DMSO, δ (ppm)): δ 148.00 (s, $C_9$), 147.62 (s, $C_2$), 145.48 (s, $C_3$), 125.42 (s, $C_6$), 123.58 (s, $C_4$), 123.02 (s, $C_5$), 107.21 (s, $C_1$), 56.12 (s, $C_7$).

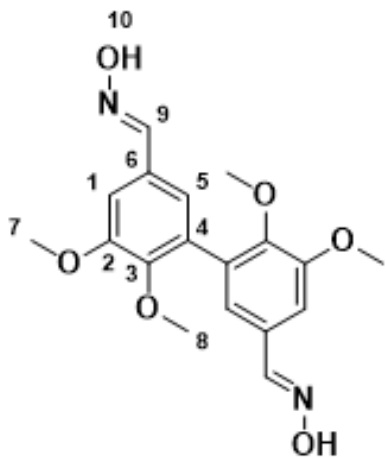

$^1$H NMR (400MHz, DMSO, δ (ppm)): δ 11.58 (s, $H_{10}$), 8.10 (s, $H_9$), 7.30 (d, $H_1$), 6.98 (d, $H_5$), 3.87 (s, $H_7$), 3.56 (s, $H_8$).

$^{13}$C NMR (400MHz, DMSO, δ (ppm)): δ 152.66 (s, $C_2$), 147.81 (s, $C_9$), 147.25 (s, $C_3$), 131.87 (s, $C_6$), 128.69 (s, $C_4$), 121.69 (s, $C_5$), 108.78 (s, $C_1$), 59.88 (s, $C_8$), 55.6 (s, $C_7$).

### Procedure for reduction of oxime

1 g of methylated divanillyloxime (2,7 mmol) and 1 mL of nickel Raney (slurry) were added in 30 mL of ethanol. The mixture was set into pressurised reactor with 10 bars of dihydrogen. After 15 h at 70 °C, the mixture was filtered and ethanol was removed under vacuum. The resulting product is extracted with dichloromethane and washed with water. Dichloromethane was removed from the organic phase under vacuum. Yield: 70%

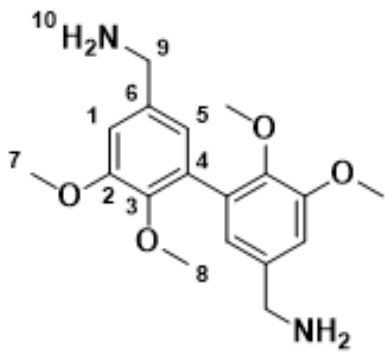

$^{1}$H NMR (400MHz, DMSO, δ (ppm)): δ 7.04 (m, $H_5$), 6.69 (m, $H_1$), 3.79 (m, $H_8$), 3.63 (s, $H_7$), 3.48 (m, $H_9$).

$^{13}$C NMR (400MHz, DMSO, δ (ppm)): δ 151.75 (s, $C_2$), 144.46 (s, $C_3$), 136.04 (s, $C_6$), 132.36 (s, $C_4$), 121.78 (s, $C_5$), 111.41 (s, $C_1$), 60.05 (s, $C_8$), 55.65 (s, $C_7$), 51.62 (s, $C_9$).

### Procedure for hydroxamic acid synthesis from ester

1,4 g of hydroxylamine hydrochloride were solubilsed in 10 mL of MeOH and 2,3g of potassium hydroxide were dissolved in 10 mL of MeOH. Both preparations were cooled down in ice, and the alkali solution was added to the hydroxylamine solution under stirring. The precipitated salts were removed by filtration and the filtrate was added to 2g of methylated diester. Additional potassium hydroxide was added to increase the pH>10 and the mixture was stirred during 12 hours at room temperature. An aqueous solution of HCl (2M) was then added to the mixture and precipitation occurred. The solid was filtered off and washed with water.

### Procedure for hydroxamic acid synthesis from oxime

1 g of methylated divanillyloxime (3 mmol) and 2,3 g of HTIB (6 mmol) were solubilised in 2 mL of DMSO and stirred at 90°C. After 24h, 0,8 g of sodium hydroxide was added to the mixture and stirred during 96h. Finally, the reaction mixture was cooled down to room temperature and 5 mL of aqueous solution of HCl (5M) was added. Precipitation occurred and the solid was filtered off and washed with water.

### Procedure for acyl azide synthesis

3 mmol of methylated divanillic acid were solubilized into mixture of 15mL of THF and 5 mL of water. The reaction mixture was cooled to 0 °C and 2,4 mL of triethylamine in 4 mL of THF was added drop-wise. 1,8 mL of ethylchloroformate was then added to the mixture and stirred during 2h at 0 °C. A solution of sodium azide (1,2g in 4 mL of water) was added drop-wise into the reaction mixture and stirred at 0 °C for 2 h and then at room temperature for 8 h. Cold water was added gradually to the reaction mixture to precipitate the solid. The precipitate was filtered off. The product was then dissolved in DCM, washed with water and dried over anhydrous magnesium sulfate. Dichloromethane was removed from the organic phase under vacuum. Yield: 60%

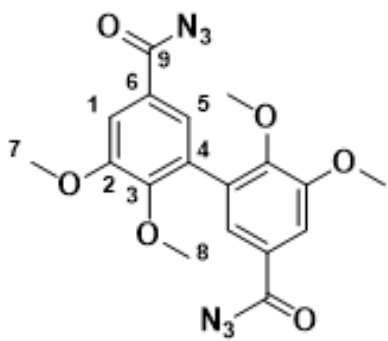

$^1$H NMR (400 MHz, CDCl3, δ (ppm)): δ 7.61 (d, $H_1$), 7.55 (d, $H_5$), 3.96 (s, $H_7$), 3.74 (s, $H_8$).

$^{13}$C NMR (400 MHz, CDCl3, δ (ppm)): δ 171.85 (s, $C_9$), 152.79 (s, $C_3$), 152.39 (s, $C_2$), 131.69 (s, $C_6$), 125.81 (s, $C_4$), 125.23 (s, $C_5$), 112.78 (s, $C_1$), 61.10 (s, $C_8$), 56.22 (s, $C_7$).

### Procedure for isocyanate synthesis

Into a Schlenk tube under nitrogen atmosphere, 0.5 mmol of diazide were solubilised in 3 mL of dry toluene and stirred. The reaction mixture was heated at 80 °C for 8 h. The toluene was removed under reduced pressure at 60 °C and white oily compound was obtained. Yield: 80%

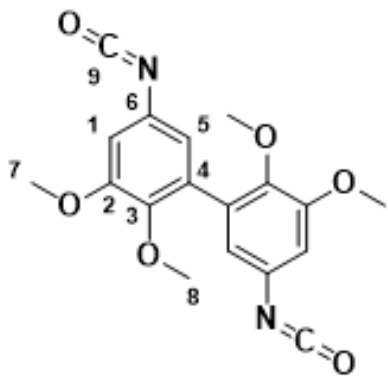

$^1$H NMR (400 MHz, CDCl3, δ (ppm)): δ 6.65 (d, $H_1$), 6.58 (d, $H_5$), 3.88 (s, $H_7$), 3.64 (s, $H_8$).

$^{13}$C NMR (400 MHz, CDCl3, δ (ppm)): δ 153.41 (s, $C_3$), 144.82 (s, $C_2$), 132.46 (s, $C_6$), 128.72 (s, $C_4$), 124.71 (s, $C_9$), 118.91 (s, $C_5$), 108.86 (s, $C_1$), 60.99 (s, $C_8$), 56.13 (s, $C_7$).

### *Procedure for isocyanate hydrolysis*

3 mmol of a potassium hydroxide solution were added to 0.75 mmol of di-isocyanate in solution in toluene. The mixture was stirred 12h at 80 °C. Toluene was removed under reduced pressure. The product was solubilised in ethyl acetate and washed with water. Ethyl acetate was removed from the organic phase under vacuum. Yield: <10%

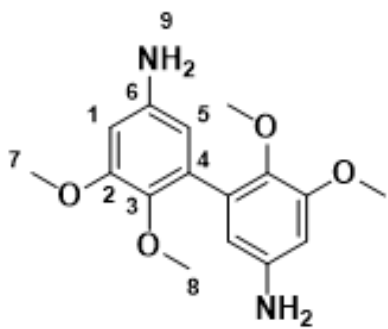

$^1$H NMR (400 MHz, DMSO, δ (ppm)): δ 6.23 (d, $H_1$), 5.90 (d, $H_5$), 4.79 (s, $H_9$), 3.72 (s, $H_7$), 3.38 (s, $H_8$).

$^{13}$C NMR (400 MHz, CDCl3, δ (ppm)): δ 152.47 (s, $C_3$), 144.22 (s, $C_2$), 136.90 (s, $C_6$), 133.35 (s, $C_4$), 107.48 (s, $C_5$), 98.42 (s, $C_1$), 59.97 (s, $C_8$), 55.18 (s, $C_7$).

### *Procedure for epoxy resin*

Epoxy monomers were mixed vigorously with stoichiometric amount (*r*=1) of IPDA, DDS, MDVA or DMAN. The mixture was then placed into an aluminium DSC pan.

## Methods

### Nuclear Magnetic Resonnance (NMR)

All NMR experiments were performed at 298 K on a Bruker Avance 400 spectrometer operating at 400 MHz. $CDCl_3$ and DMSO-d6 were used as deuterated solvent depending on the sample and specified in the legends of spectra.

### Flash chromatography

Flash chromatography was performed on a Grace Reveleris apparatus, employing silica cartridges from Grace and a dichloromethane/methanol gradient solvent equipped with ELSD and UV detectors at 254 and 280 nm.

### Differential Scanning Calorimetry (DSC)

Differential Scanning Calorimetry (DSC) measurements were performed on DSC Q100 (TA Instruments). The sample was heated at a rate of 10 °C.min$^{-1}$. Consecutive cooling and second heating run were also performed at 10 °C.min$^{-1}$. The glass transition temperatures and melting points were calculated from the second heating run.

### Dynamic Mechanical Analysis (DMA

Dynamic Mechanical Analysis (DMA) measurements were performed on DMA-RSA3 system from TA instruments. The three point bending sample (width = 2 mm; thickness = 2 mm and length of fixed section = 10 mm) was heated from 25 °C to 350 °C at a heating rate of 5 °C.min$^{-1}$. The measurements were performed in a three-point bending mode at a frequency of 1 Hz, an initial static force of 0.5 N and a strain sweep of 0.01 %.

### Thermogravimetric analyses (TGA)

Thermogravimetric analyses (TGA) were performed on TGA-Q50 system from TA instruments at a heating rate of 10 °C.min$^{-1}$ from room temperature to 950 °C. The analyses were investigated under air and nitrogen atmosphere with platinum pans.

### Fourier Transformed Infra-Red-Attenuated Total Reflection (FTIR-ATR)

Infrare spectra were performed on a Bruker VERTEX 70 spectrometer, equipped with diamond crystal (GladiATR PIKE technologies) for attenuated total reflection mode. The spectra were acquired from 400 to 4000 cm$^{-1}$ at room temperature using 32 scans at a resolution of 4 cm$^{-1}$.

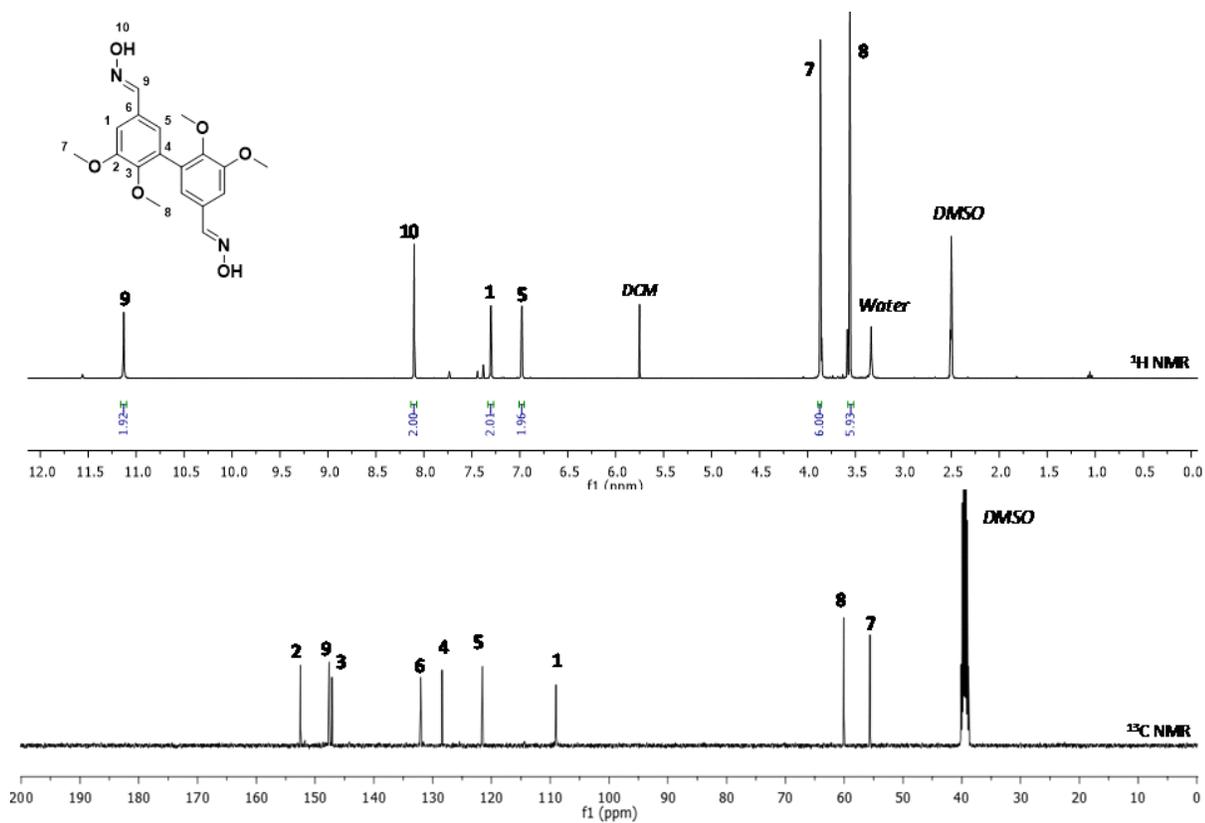

Figure S1 : $^1$H and $^{13}$C NMR spectra of methylated divanillyloxime in DMSO-d6

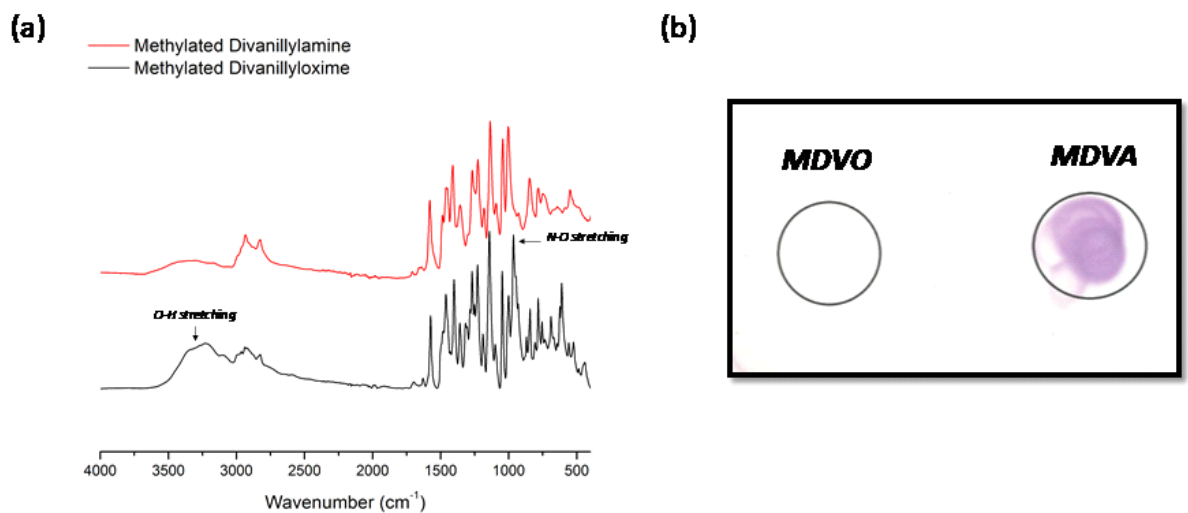

Figure S2 : (a) FTIR spectra of MDVA and MDVO, (b) Ninhydrin test results of MDVO and MDVA on gel silica plate

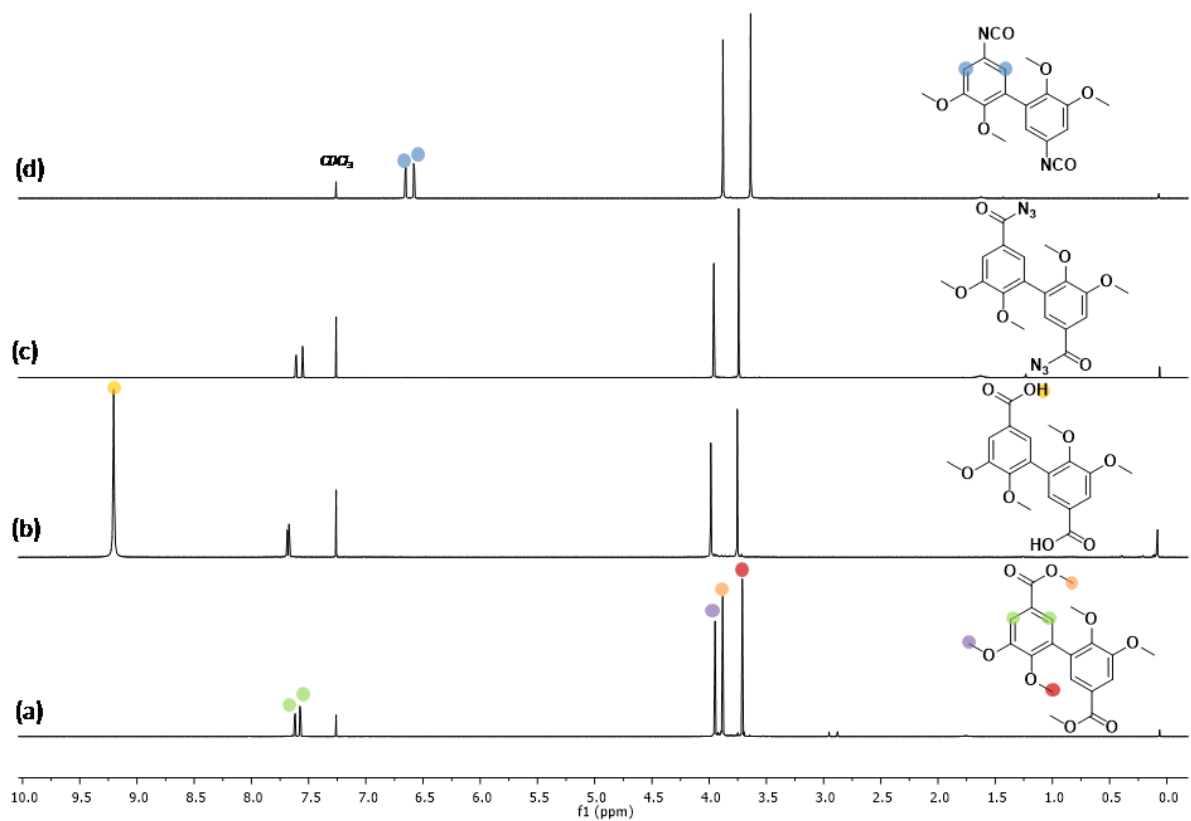

Figure S3 : $^1$H NMR spectra of (a) methylated diester, (b) methylated divanillic acid, (c) methylated diacyl azide and (d) methylated di-isocyanate of vanillin in CDCl$_3$

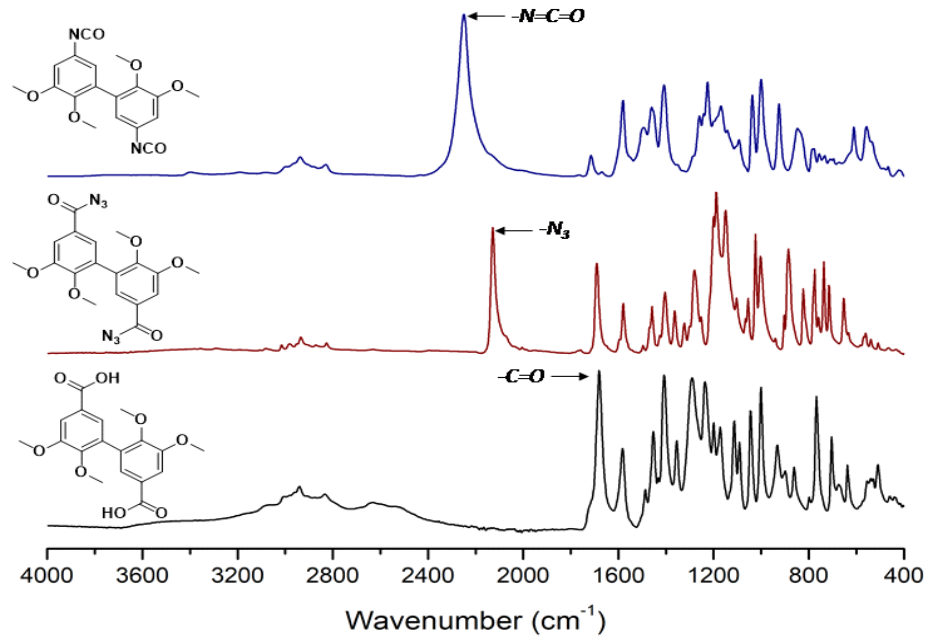

Figure S4 : FTIR spectra of methylated divanillic acid, di-acyl azide and di-isocyanate

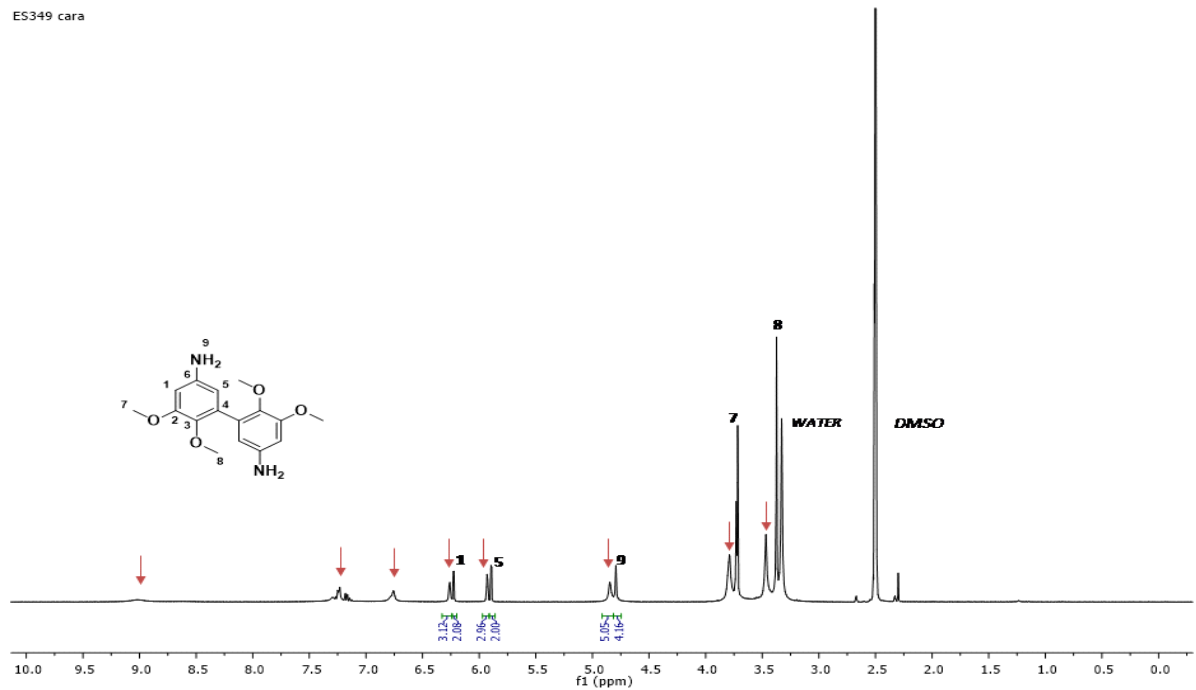

Figure S5 : $^1$H spectrum of 3,4-dimethoxydianiline before purification in DMSO-d6

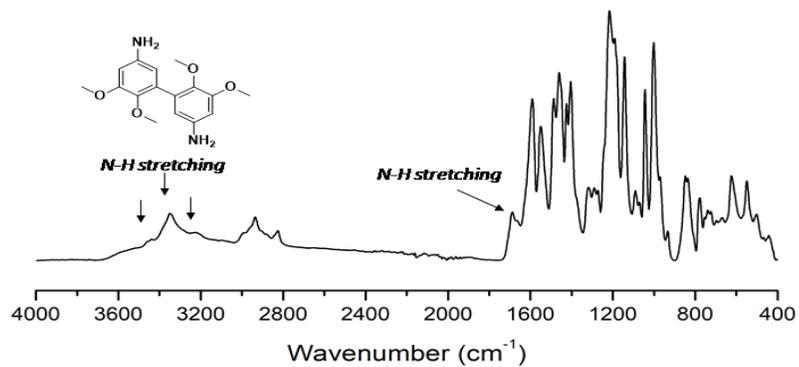

Figure S6 : FTIR spectrum of 3,4-dimethoxydianiline

| Epoxy/hardener | T_Onset (°C) | T_Exotherm (°C) | ΔH (J.g⁻¹) | Tg (°C) | Char₉₀₀ (%) | |
|---|---|---|---|---|---|---|
| | $T_{Onset}$ (°C) | $T_{Exotherm}$ (°C) | $\Delta H$ (J.g$^{-1}$) | $T_g$ (°C) | Char$_{900}$ (%) | *No Tg was clearly |
| DGEBA/IPDA | 73 | 111 | 430 | 152 | 8 | |
| DGEBA/MDVA | 96 | 142 | 105 | * | nd | |

*observed by DSC*

Table  S1 : Thermomechanical properties of DGEBA cured with IPDA and MDVA

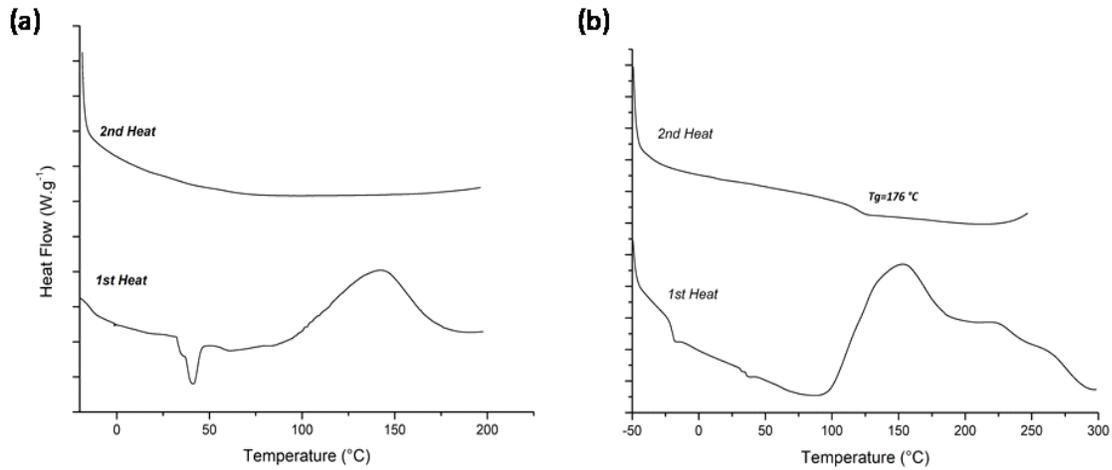

Figure S7 : DSC thermograms of (a) DGEBA/MDVA and (b) DGEBA/DMAN

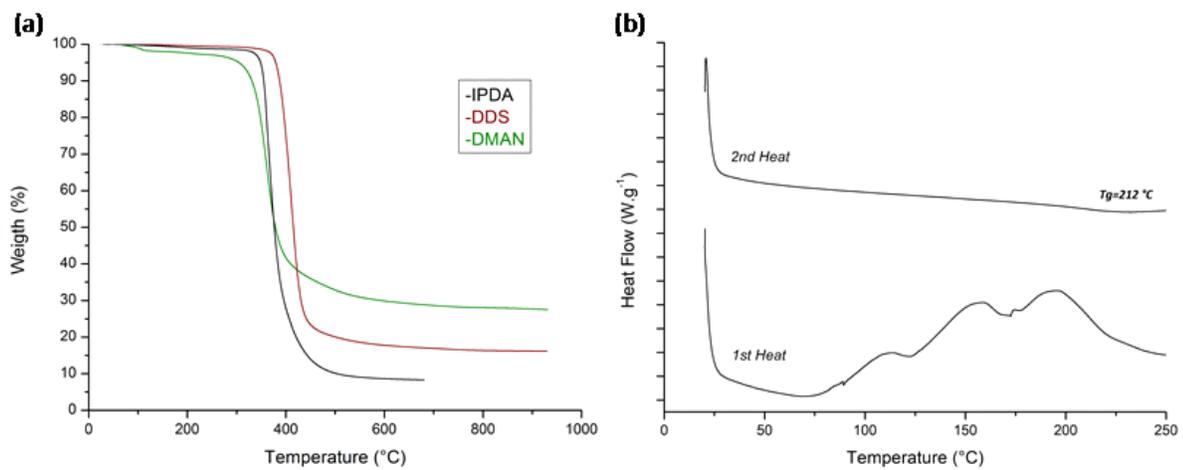

Figure S8 : (a) TGA thermograms of DGEBA cured with IPDA, DDS and DMAN, (b) DSC thermograms of TetraGEDVA cured with DMAN